# A Simplified Phase Model for Oscillator Based Computing

[1]Yan Fang, [2]Victor. V. Yashin, [3]Donald M. Chiarulli, [1]Steven P. Levitan
[1]Department of Electrical and Computer Engineering, [2]Department of Chemical Engineering, [3]Department of Computer Science
University of Pittsburgh
Pittsburgh, PA 15260

*Abstract*—Building oscillator based computing systems with emerging nano-device technologies has become a promising solution for unconventional computing tasks like computer vision and pattern recognition. However, simulation and analysis of these systems is both time and compute intensive due to the nonlinearity of new devices and the complex behavior of coupled oscillators. In order to speed up the simulation of coupled oscillator systems, we propose a simplified phase model to perform phase and frequency synchronization prediction based on a synthesis of earlier models. Our model can predict the frequency locking behavior with several orders of magnitude speedup compared to direct evaluation, enabling the effective and efficient simulation of the large numbers of oscillators required for practical computing systems.

## 1. INTRODUCTION

Pursuing high-density, low-power, high-speed computing systems for the post-CMOS era drives researchers to exploit the potential of emerging nano-device technologies. Based on recent advances in emerging nano-devices such as spin torque oscillators (STO) [1][2], resonate body transistor oscillators (RBO) [3], and vanadium oxide oscillators (VO2) [4], systems built from coupled nano-oscillators have become promising candidates for next generation computing structures used in intelligent information processing [5]. Inspired by the interaction between neural oscillations that occurs at different scales in biological systems, Hoppensteadt and Izhikevich developed an associative memory model of coupled oscillators by using phase locked loops and provide examples about how this dynamic system performs recognition by forming attractor basins at the minima of a Lyapunov energy function [6].

The essential idea of utilizing coupled oscillator systems to perform computation lays in the energy transfer in a dynamical system. Initialized with input information, a number of oscillators interact and exchange energy with each other, making the whole dynamical system converge from a perturbed state to a stable state. This process brings several advantages: First, it provides a high level multi-dimension norm, like the Euclidean distance, between sets of input vectors. Compared to the Hamming distance computed with exclusive-OR Boolean operations, the oscillator clusters are capable of processing matching operations in parallel and giving a robust pattern match computation. The high frequency of oscillations of new devices like STOs means that the systems can converge extremely fast. More importantly, this computing structure is very suitable for large high dimension data sets due to its high scalability and the degree of match that spans all of the dimensions of a input vector without any arithmetic calculations. Recently, research on these systems has been conducted not only from the perspective of devices and circuits [7][2], but also at the level of algorithms and architectures and present structures that address the scalability of oscillator computing models [8].

In order to design and build these systems, we need a good understanding of the behavior of coupled oscillators, including the synchronization and de-synchronization between oscillators, the prediction of the coupled oscillators' frequencies, and their relation to a "degree of match" function in pattern matching. Unfortunately, accurate simulation and analysis from the device level to the architecture level is difficult and computationally expensive for current EDA/CAD tools. This is because of both the complexity of the nano-oscillators' device models and the short time scales necessary to capture the coupling dynamics. Thus, circuit and system designers cannot model systems at very large scales, while algorithm designers and architects can only use approximate distance norms instead of using the oscillators' true behaviors. To address this problem, we focus on modeling the ensemble behavior of coupled oscillators.

Winfree [9] and Kuramoto [10] made fundamental contributions to the understanding of these systems. Hoppensteadt and Izhikevich studied weakly coupled oscillator networks and unified previous work by using the idea of the phase resetting curve (PRC) [11]. Recently, Roychowdhury proposed a nonlinear phase model based on a perturbation projection vector (PPV) to study the perturbations due to noise of electronic oscillators [12] and used this to model coupled oscillators. These models were widely used in modeling the injection locking of oscillators to external signals [13]. Mafizzoni developed this model one step further and analyzed weakly coupled oscillators from the perspective of multi-frequency analysis [14]. The contribution of this work is the synthesis of oscillator phase model with the PPV model to provide a new model that simplifies the analysis of large systems.

The rest of this paper is organized as follows. First, we study different phase models for coupled oscillators and show the equivalency between them. Then, we combine ideas from these models to abstract a simplified model for oscillator based pattern matching operations. Finally, we provide comparative simulation results and conclude with observations about the effectiveness of our model and future work.

## 2. PHASE MODELS

In this section, we review phase models proposed in previous work, including the PPV model [12], Winfree and Malkin's approach summarized by Izhikevich [11]. By discussing the relation between these models and combining their advantages, we develop our new model.

## 2.1 PPV Model

We start the introduction of the model with the PPV model, since it is derived from the demand for a theoretical analysis of oscillator circuits perturbed by noise and therefore is easier to understand from an electrical engineering perspective compared to the other models.

The PPV model [12] starts with the general differential equation of an oscillator,

$$\dot{x}(t) = f(x(t)) + P(t) \quad (1)$$

where x(t) is the vector of oscillation states and ẋ(t) is their derivatives. In real circuits, these states are usually the voltages or currents of nodes. P(t) represents an external perturbation on the oscillation, which can be noise, signal injection, or a coupling term from other oscillators, as in our case. $f$ is the nonlinear function that describes the oscillation and $t$ is time. Then, the response solution from this equation can be written as,

$$x_c(t) = x_s(t + \alpha(t)) \quad (2)$$

where $x_s(t)$ is the oscillator's natural response without any perturbation, namely $P(t) = 0$, while $x_c(t)$ is the response with perturbation. $\alpha(t)$ represents the time shift of phase that is caused by the perturbation. Hence equation (2) reveals the phase relation between the natural response and perturbed response of the oscillator. According to the PPV model, $\alpha(t)$ can be obtained by solving the equation,

$$\dot{\alpha}(t) = \Gamma(t + \alpha(t))C(t) \quad (3)$$

where $\Gamma(t)$ is the perturbation projection vector (PPV) that describes the perturbation response of the oscillators. The PPV can be thought of as the time-varying sensitivity of the induced time shift to the given injected perturbation [13]. The theoretical derivation and proof of the PPV method can be found in [12]. The PPV $\Gamma(t)$ is usually a periodic function that can be obtained either numerically or analytically [12][13]. After acquiring $\Gamma(t)$, we know the time shift of phase $\alpha(t)$ from the solution of equation (3) and then solve equation (2).

This model actually does not predict the "phase change" but a time shift function of oscillation response. Thus, it is difficult for us to use this model to directly predict the frequency shift in a coupled oscillator system. The next models we introduce provide us further insight.

## 2.2 Izhikevich's Model

In [11], the phase model of weakly coupled nonlinear oscillators is explored from an abstract view, by unifying some earlier models. Differing from the PPV model, this model assumes all the oscillators share the same free-running frequency, so the "phase" of nonlinear coupled oscillators can be normalized and defined as,

$$\theta(t) = t + \varphi(t) \quad (4)$$

Taking the derivative from both sides, we have

$$\dot{\theta}(t) = 1 + \dot{\varphi}(t) \quad (5)$$

In equations (4)(5), $\theta(t)$ is the defined phase, a periodic function with period T = 1. $\varphi(t)$ is called the phase deviation, caused by the coupling from other oscillators. We can notice that when there is no coupling term, the phase term is simply time, $t$, and the free-running frequency is normalized to 1. The derivative $\dot{\varphi}(t)$ represents the change of phase deviation, namely the frequency shift, due to the coupling effect.

In order to map this model to various nonlinear oscillators, the key point lies in the phase deviation $\varphi(t)$, described by:

$$\dot{\varphi}(t) = Q(\theta(t))P(t) \quad (6)$$

which has a similar form to equation (3). $P(t)$ is the same external injection signal to the oscillators (i.e., the coupling term). The function $Q(\theta)$ is called the phase response curve or phase resetting curve (PRC), representing how sensitive the phase deviation is in response to P(t) at a specific phase $\theta(t)$. Thus (5) can also be written as,

$$\dot{\theta}(t) = 1 + Q(\theta(t))P(t) \quad (7)$$

The mechanism of phase and frequency of coupled oscillators revealed by this equation was discovered multiple times in the early studies of oscillator phase models. But the researchers named it PRC in different ways and exploited different methods to derive and utilize it.

A theorem first proposed by Malkin in 1949 [17] and later abstracted by Hoppenstead in 1997 indicates that $Q(\theta)$ can be computed by solving the linear adjoint equation,

$$\dot{Q}(t) = -\mathcal{J}[f(x(t))]^T Q \quad (8)$$

where $\mathcal{J}[f(x(t))]^T$ is the transposed Jacobian matrix of the oscillation function f in (1). This theorem is identical to Kuramoto's approach in 1984, where the gradient of phase plays the role of PRC.

In Winfree's work (1967), PRC was experimentally approached by applying a pulse stimuli with amplitude A. Then a function called the linear response or sensitivity function $Z(\theta)$ was measured by observing the phase shift caused by the stimuli, and PRC is $Z(\theta)$ divided by amplitude $A$.

$$PRC(\theta) \approx \frac{Z(\theta)}{A} \quad (9)$$

## 2.3 Our Simplified Model

From the previous models discussed above, if we pair equations (1)(4), (2)(5), (3)(6) and compare them, we can notice that they have the same pattern because the intrinsic method behind these models is the same, which is to quantify how the oscillation is affected by the external perturbation. However, these models use different methods to calculate this term, either an analytical derivation or a numerical measurement. In addition, the PPV model studies the oscillation variables while Izhikevich's model focuses on the "phase" of nonlinear oscillators.

From the view of solving the practical problem we are addressing, the pattern matching operation is performed by frequency shifting or frequency locking of oscillators caused by coupling. The elements of the pattern vectors are represented by frequencies and the degree of match is evaluated by how well the oscillators synchronize. Thus, for the purpose of predicting the frequencies of coupled oscillators, we introduce the phase definition idea from Izhikevich's model into the PPV model.

We assume we have n oscillators with different frequencies: $\omega_0, \omega_1, \omega_2 \ldots \omega_{n-1}$. Since the equation (4) and (5) require that oscillators run at the natural frequencies normalized to 1, we scale these frequencies to $1, \lambda_1, \lambda_2 \ldots \lambda_{n-1}$, where $\lambda_i = \frac{\omega_i}{\omega_0} = \frac{T_0}{T_i}$. So for an arbitrary oscillator i, in the PPV model, equation (2) can be changed into the phase form similar to (4):

$$\theta_i(t) = \lambda_i(t + \alpha(t)) = \lambda_i t + \varphi_i(t), \quad (10)$$

with

$$\varphi_i(t) = \lambda_i(\alpha(t)), \quad (11)$$

(10) indicates the relation of phase deviation in Izhikevich's model and time shift of phase in the PPV model. Taking the derivative of (9) we get,

$$\dot{\theta}_i(t) = \lambda_i + \lambda_i \dot{\alpha}(t), \quad (12)$$

substituting with (3) we have,

$$\dot{\theta}_i(t) = \lambda_i + \lambda_i \Gamma(\theta_i(t)) P_i(t) \quad (13)$$

where $\Gamma(\theta_i(t))$ is still the PPV in (3) but determined by the phase, instead of time. This equation transfers the PPV from a time domain to the phase domain and replaces the simulation time span into the number of oscillation cycles. $C_i(t)$ is the coupling term in this model, defined as:

$$C_i(t) = \sum_{j=0}^{n} g_{ij} x_s(\theta_j(t)) \quad (14)$$

where $g_{ij}$ is the coupling coefficient and $j$ is the index of other oscillators in the system. Therefore, solving (12) can provide us the frequency and phase response of a coupled oscillator system.

In our simplified model, the PPV function is also equivalent to the PRC function obtained from other methods. In the next section we give examples of these methods. We note that it is useful to have several methods available because some methods might prove inaccurate or fail to converge for specific nonlinear oscillator systems.

## 3. OSCILLATOR EXAMPLES

In this section we use the nonlinear ring oscillators as an example and demonstrate three different methods to obtain their PPV/PRC function.

A simple ring oscillator consists of three inverters and RC circuits, shown in Figure 1(a) with its analytical model given in [13]. The voltage derivatives of the three nodes are,

$$\dot{v}_i(t) = \frac{f(v_{i^-}(t)) - v_i(t)}{RC}, i = 1,2,3 \quad (15)$$

where $i^-$ is the previous node of $i$ and f(v(t)) is the simplified response of an inverter:

$$f(v) = \begin{cases} +1, & if\ v > 0 \\ -1, otherwise \end{cases} \quad (16)$$

By normalizing the standard frequency and period to 1, we can write the voltage state response of three nodes from [18] into,

$$v_1(t) = \begin{cases} 1 - \psi e^{-\gamma t}, & 0 \le t \le 1/2 \\ -1 + \psi e^{-\gamma\left(t-\frac{1}{2}\right)}, & 1/2 \le t \le 1 \end{cases}$$

$$v_2(t) = v_1\left(t - \frac{2}{3}\right), v_3(t) = v_1\left(t - \frac{1}{3}\right) \quad (17)$$

where $\psi = \frac{1+\sqrt{5}}{2}, \gamma = 6\ln(\psi)$. Because the frequency here is 1, $RC = 1/\gamma$. Similarly, the PPV equation in [18] can be analytically solved as,

$$\Gamma_1(t) = \Gamma_3\left(t - \frac{2}{3}\right)$$

$$\Gamma_2(t) = \Gamma_3\left(t - \frac{1}{3}\right)$$

$$\Gamma_3(t) = \gamma^{-1} \frac{1+\psi^3}{4-2\psi^3} \left(\psi + 2\left[-u(t) + (-1 + 2\psi^{-1})u\left(t - \frac{1}{2}\right)\right]\right) e^{\gamma t} \quad (18)$$

Figure 1(b) shows the waveform of oscillation response and PPV of node v3. From this plot we see that the PPV function for ring oscillators is periodic but not linear.

Sometimes, it is difficult to obtain the state response and PPV function directly from the ODE. For these cases we can obtain the corresponding PRC by applying Malkin's approach numerically. Solving equation (8) is actually very similar to the analytically derivation of PPV in [15]. However, when the analytical method does not work, we can use a technique called backward integration to obtain the Jacobian matrix [13]. Figure 2 shows the results of PRC from this method with a backwards integral of 4 cycles. The PRC in the first cycle is the one used.

In a few cases when a system's Jacobian matrix does not exist, Winfree's approach could be the only choice, especially for those nonlinear oscillators with complex mathematical models, even though this method is "experimental" and tends to be inaccurate. As an example, we apply Winfree's method for the ring oscillator by adding a pulse stimulus with small amplitude and measuring the phase resetting curve step by step. Figure 3 illustrates the PRC generated by Winfree's method with different stimuli amplitudes. The glitches in the curve show the problems of this approach.

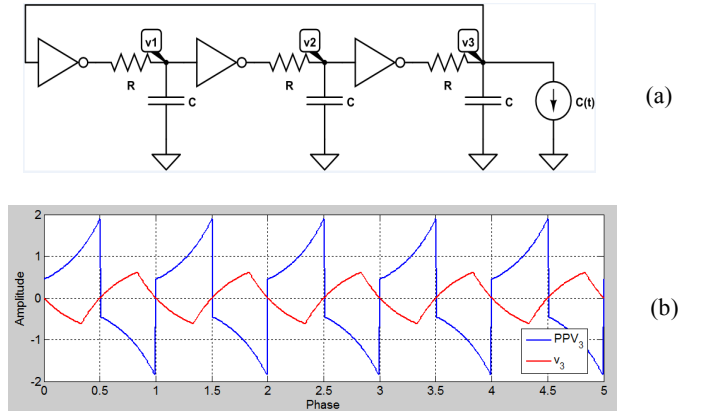

**Figure 1. Ring oscillator model. (a) Simple schematic model; (b) Output waveform (red) and (PPV) (blue) at node v3.**

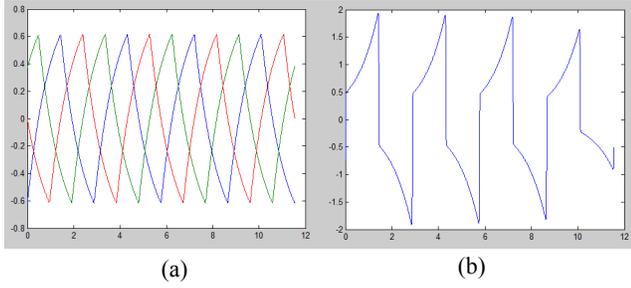

(a) (b)

**Figure 2. Malkin's approach for PRC. (x-axis: phase, y-axis: amplitude) (a) Oscillation waveform of voltage at three nodes; (b) PRC obtained from backward integral.**

However, it is worth noting that the PRC amplitude here is proportional to the stimuli amplitude by factor of 2, which not only corresponds to equation (9) but also fits the PPV/PRC amplitude of the previous two methods in Figure 1(b) and Figure 2 (b). These three examples for phase deviation of the ring oscillator indicate that PPV and PRC functions are identical to each other and enhance the foundation of our model.

## 4. EXPERIMENTS AND SIMULATION
### 4.1 System Configuration

In this section, we apply our simplified phase model to coupled oscillator systems and analyze their synchronization behavior. We also compare the performance and efficiency of the models obtained by the different approaches as well as accuracy and speedup compared to the direct simulation of the oscillator systems.

Figure 4 gives a circuit example of our coupled oscillator system for pattern matching. For ring oscillators, the frequency of each oscillator is adjusted by two input control voltages and the coupling node is the output node of the third inverter. These oscillators are coupled to each other through a common node. The coupling component could be resistor or capacitor, chosen by the circuit designer, where larger resistors would give weaker coupling strength. The detection circuits that read out the degree of synchronization can be found in [8].

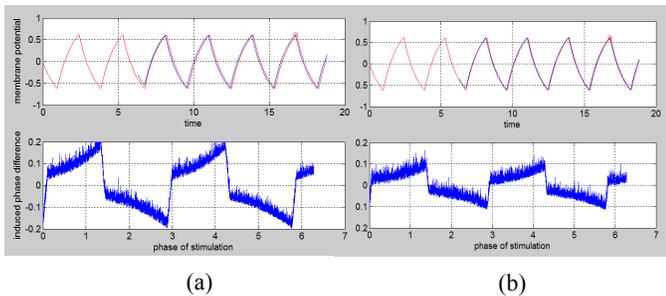

(a) (b)

**Figure 3. Winfree's approach for PRC, (a) Stimuli Amplitude=0.1, PRC Amplitude=0.2; (b) Stimuli Amplitude=0.05, PRC Amplitude=0.1.**

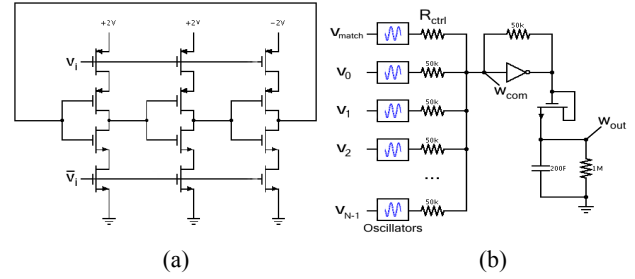

(a) (b)

**Figure 4. Coupled oscillator system. (a) Voltage controlled ring oscillator; (b) Oscillator coupled together through one common node with synchronization detection circuits.**

As we discussed in the previous section, such a weakly coupled system can be simply described by equations (13) and (14). In this structure the coupling strengths between oscillators are identical. Thus, for each oscillator (with coupling at node v3 in Figure 1) we have,

$$\dot{\theta}_i = \lambda_i + \varepsilon \lambda_i \, \Gamma_{node3}(\theta_i) \sum_{j=0}^{n} v_{node3}(\theta_j), i,j \in [1,n] \quad (19)$$

where ε is the coupling coefficient.

### 4.2 Oscillator Behavior Analysis

We start with a three oscillator system ($n = 3$). It is convenient for us to predict the final frequency of each oscillator by solving (19) numerically. If $\dot{\theta}_1 = \dot{\theta}_2 = \dot{\theta}_3$, the system is synchronized and frequency locked. We use Matlab to run a simulation of equation (11). Figure 5 illustrates an example of locking and non-locking systems. For these examples, we set the initial frequencies to be: $\lambda = [1, 0.95, 1.05]$ but use two different ε, 0.2 and 0.4. In the left plot, $\varepsilon = 0.4$ and the final frequencies are [0.8717, 0.8717, 0.8717]; in the right plot, $\varepsilon = 0.2$ and the final frequencies are [0.9644, 0.9298, 1.0263].

From (19) and these simulation results, it is worth noting that the conditions for frequency locking are determined by the coupling coefficient ε and the scaling ratio between each oscillators' free running frequency $\lambda_i$, not the absolute value of the frequencies. This interesting phenomenon is important for the design of future oscillator based computing systems. It implies that devices with high frequencies can provide wider bandwidth for information coding. Also, even when the oscillators fail to lock with each other, their frequencies are pulled closer to each other. It is based on this observation that we say that the degree of synchronization can provide a metric to measure the distance or similarity of each oscillator's initial frequency (and thus input) as discussed below.

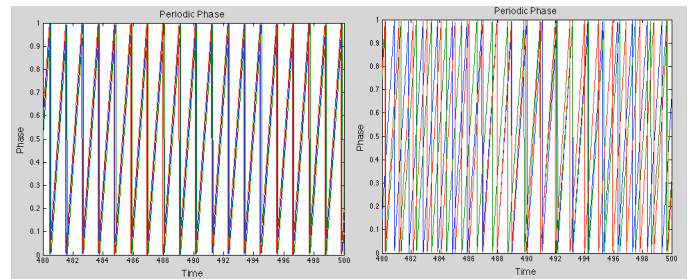

**Figure 5. Simulations of three coupled oscillators with natural frequencies λ = [1.0, 0.95, 1.05]. Left: ε=0.4, Right: ε=0.2.**

To verify this point, we run the simulation multiple times by fixing the first oscillator's frequency and sweeping the frequency of the other two from 0.8 to 1.2. We use $1 - \{(f_2 - f_1)^2 + (f_3 - f_1)^2\}$ as the function to evaluate how well the oscillators synchronize. While there are many possible metrics, like standard deviation of frequencies, that can analyze the behavior of coupled oscillators, here we choose the inverted Euclidean distance. We use this function to capture the synchronization of oscillators in order to show the difference between the phase model and direct simulations and thus to verify the capability of our phase model in simulation of oscillator based computing. Note that $f$ here is the coupled scaled frequency. In this test, we use three different methods to obtain the PPV/PRC for our model, as we did in section 3. In addition, we directly simulate the coupled oscillator without using any phase model as a performance standard. Figure 6 demonstrates all the cases with 3D plots of the degree of synchronization for the two cases of $\varepsilon = 0.4$ and $\varepsilon = 0.2$. When the coupling strength is weak, the surface is smooth and the initial frequencies are easier to differentiate, while stronger coupling gives us more nonlinear features and a wider locking range. The flat area on the top of the surface indicates the highest degree of synchronization, giving us the frequency locking range for the simulation sweeps. Hence, a very strong coupling system may lack differentiation for pattern matching or nearest neighbor searching. But for clustering operations like image segmentation, stronger coupling strength can provide better resistance to noise.

Due to the fast simulation speed, our model is also very suitable for simulating systems with large numbers of coupled oscillators. To understand how the number of oscillators can influence the synchronization, we run the simulations with the same two dimensional frequency sweeping but different numbers of oscillators. Since we cannot show a plot of higher dimensional frequency sweeping, we keep the frequencies of all but two of the oscillators fixed to 1 and sweep the last two. Figure 7 shows the results for n=3, 4, 8, and 16. In these simulations, we use the analytical PPV as the PPV/PRC function.

From these results we can notice that larger numbers of oscillators with the same frequency gives a larger frequency locking range. Namely, the effective coupling strength to those oscillators with different frequencies becomes stronger because the system's stable state is close to the state of the majority of the oscillators. This could either be an advantage or disadvantage in the design of oscillator based systems, depending on the application and computation required.

### 4.3 Performance and Speedup

For performance comparisons of our phase model, we calculate the root mean square error between the direct simulation of the oscillators in each case of Figure 6 and the analytical PPV. As we can see in Table 1, the error is relatively small compared with the absolute value of degree of synchronization. This shows that our phase model is compatible with different methods for PPV/PRC and robust to the variations between these methods. Compared to the direct oscillator simulation, the analytical PPV generates the smallest error while Winfree's experimental method gives the largest error, and Malkin's numerical method lies in the middle.

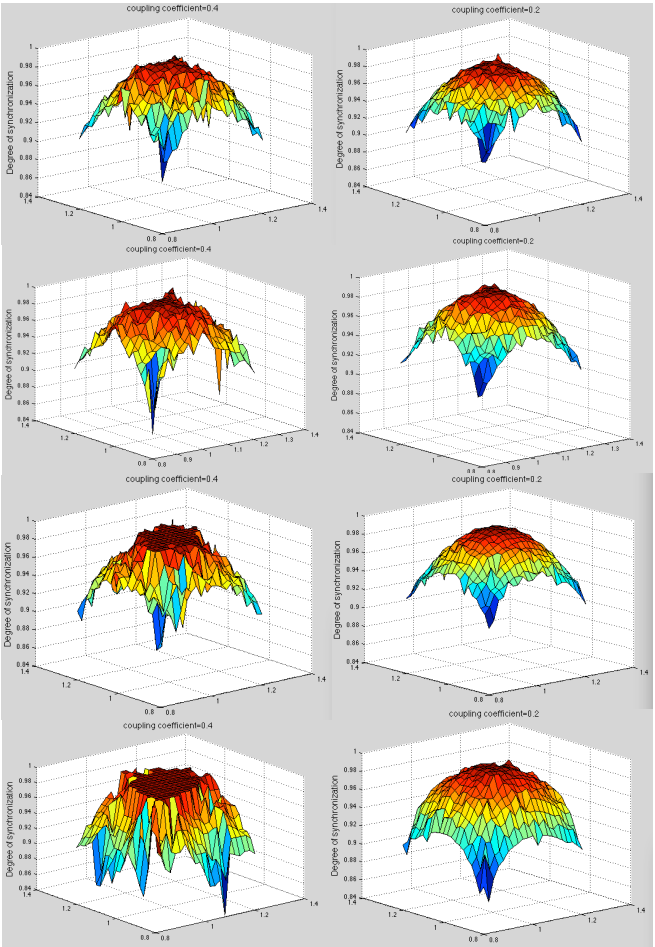

**Figure 6. Degree of synchronization as inverted Euclidean distance, initial frequencies for a range $\lambda_{2,3} = [0.8 \text{ to } 1.2]$ and a fixed frequency of $\lambda_1$ =1.0, Left column: ε=0.4, Right column: ε=0.2; Row 1: Winfree's approach, Row 2: Malkin's approach, Row 3: Analytical PPV, Row 4: Direct simulation.**

**Table 1. RMSE of simulations based on different methods**

| Comparing | Winfree's | Malkin's | Analytical PPV |
|---|---|---|---|
| **Oscillator, $\varepsilon = 0.2$** | 127e-04 | 133e-04 | 121e-04 |
| **Oscillator, $\varepsilon = 0.4$** | 276e-04 | 297e-04 | 271e-04 |
| **Analytical PPV, $\varepsilon = 0.2$** | 58e-04 | 63e-04 | / |
| **Analytical PPV, $\varepsilon = 0.4$** | 128e-04 | 142e-04 | / |

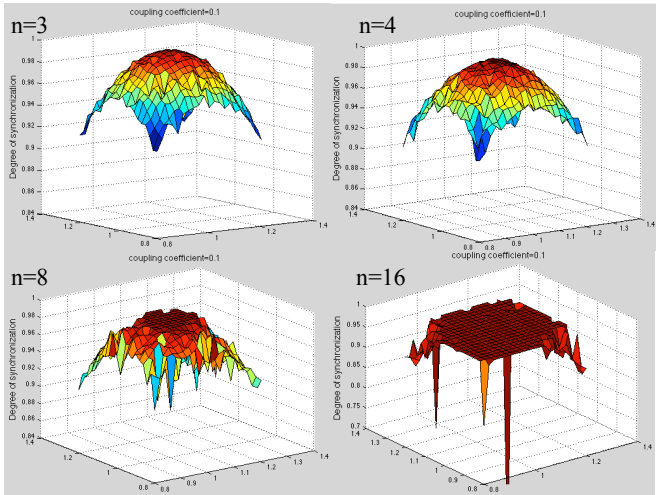

**Figure 7. Simulations of different numbers of coupled oscillators. All oscillators but two are kept at frequency 1, while the last two are swept.**

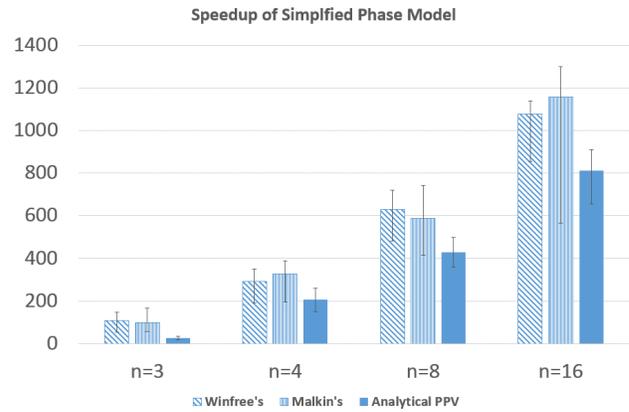

**Figure 8. Speed-up (factor) of simplified phase model over direct simulation for different PPV/PRC methods.**

We next evaluate the efficiency of our model by comparing simulation speed to the direct simulation of the oscillator network. The speedup here is defined by the ratio of real time for simulation of two methods (both in Matlab) for the same length of simulation time. Since we always initialize the systems with random phases, we run the test of each configuration 100 times and average the speedup for evaluation. The results in Figure 8 only serve as an approximation of the efficiency of our model because the simulation of coupled oscillator systems is affected by multiple factors, such as the oscillator model, initial states, and the convergence process. Nevertheless, we still observe very promising speedups from our simplified phase model, similar to [14]. The speedup comes from the fact that the differential equations of the original oscillators are nonlinear but the simplified phase models are linear equations with much simpler periodic functions. A second advantage is that in the phase domain, simulation is done by fractions of cycles, rather than time steps so that run time is frequency invariant. This is an advantage in simulation when compared to the regular PPV model where the user must optimize the time step for accurate simulation vs. performance.

## 5. CONCLUSIONS

In this paper, we introduce the problems of oscillator simulation encountered in the design of oscillator based computing systems. To address this problem, we review previous phase models and propose a reduced simple phase model. We apply our model in the analysis of ring oscillators. The results show that our model is capable of predicting the frequency and phase of coupled oscillator systems with small errors compared to direct simulation of the oscillator model. The main contribution of our model is simplifying the nonlinear equations and transferring simulation from traditional time domain into phase domain. This provides very promising simulation speedup as the size of the system increases. Furthermore, we demonstrate that this model is particularly suitable for simulation and analysis for oscillator based computing operations like pattern matching and nearest neighbor search.

## 6. ACKNOWLEDGEMENTS

This work was supported in part by the National Science Foundation under grants DMR-1344178 and CCF-1317373.